\documentclass[prb,aps,twocolumn,floatfix,10pt]{revtex4-1}
\usepackage{amssymb,graphicx,color,amsmath,bm} 
\usepackage[ bookmarks, colorlinks=true, breaklinks]{hyperref}  
\usepackage[none]{hyphenat}
\hypersetup{linkcolor=blue,citecolor=blue,filecolor=black,urlcolor=blue} 

\newcommand{\bra}{\left< }
\newcommand{\ket}{\right>}
\newcommand{\ovl}[1]{\overline{#1}}

\begin{document}

\title{Structure Factor of a Relaxor Ferroelectric}
\author{G. G. Guzm\'{a}n-Verri$^{1,2,3}$ and C. M. Varma$^{1}$}
\affiliation{$^1$Department of Physics and Astronomy, University of California, Riverside, California, 92521, USA }
\affiliation{$^2$Materials Science Division, Argonne National Laboratory, Argonne, Illinois, 60439, USA }
\affiliation{$^3$Centro de Investigaci\'{o}n en Ciencia e Ingenier\'{i}a de Materiales and Escuela de F\'{i}sica, Universidad de Costa Rica, San Jos\'{e}, 2060, Costa Rica}

\date{\today}
\begin{abstract}
We study a minimal model for a relaxor ferroelectric including dipolar interactions, 
and short-range harmonic and anharmonic forces for the critical modes 
as in the theory of pure ferroelectrics together with quenched disorder
coupled linearly to the critical modes. 
We present the simplest approximate solution of the model necessary to obtain 
the principal features of the correlation functions. 
Specifically, we calculate and compare the structure factor measured by neutron 
scattering in different characteristic regimes of temperature in the relaxor PbMg$_{1/3}$Nb$_{2/3}$O$_3$.
\end{abstract}
\maketitle

\section{Introduction}
Relaxor ferroelectrics, a typical example of which is PbMg$_{1/3}$Nb$_{2/3}$O$_3$~(PMN), a perovskite in which the Mg$^{2+}$ 
and Nb$^{5+}$ randomly occupy the octahedrally coordinated site, have a very high dielectric constant 
for a very wide region in temperature and have no ferroelectric transition except in applied electric 
fields. Cowley {\it et al.}~\cite{Cowley2011a} in a detailed review of  relaxor ferroelectrics 
have commented that although they were first synthesized over 50 years ago and their properties have been well explored, 
a satisfactory theory explaining their properties has not yet been formulated. Here, we present
 a minimal model for the canonical relaxor PMN and its simplest necessary approximate solution based on 
Onsager's results on models with dipolar interactions, the displacive transition model of pure ferroelectrics, and quenched random fields. 

The properties of the relaxors such as PMN may be summarized as follows:~\cite{Cowley2011a, Bokov2006a, Kleeman2006a, Samara2003a, Cross1987a} 
in zero external electric field they have a region of collective dielectric fluctuations which is bounded at the upper 
end by a temperature $T_B$, the Burns temperature,~\cite{Burns1983} above which the susceptibility has a Curie-Weiss 
form.~\cite{Viehland1992a} 
A broad region extending down to $T=0$ below $T_B$  is marked by a temperature $T_{\max}$ where 
the susceptibility is maximum and has a width in temperature $\Delta T_{\max}$. Both  
$T_{\max}$ and $\Delta T_{\max}$ depend on the frequency $\omega$ at which the 
dielectric susceptibility is measured.~\cite{ Smolenskii1970a, Bovtun2004a, Hlinka2006a, Bokov2012a, Glazounov1998a} 
Very significantly, neutron scattering experiments have revealed that 
the structure factor $S_{\bm q}$ has unusual temperature~\cite{Gvasaliya2005a, Hiraka2004a, Gehring2009a}, power-law~\cite{Gvasaliya2005a} 
forms and anisotropies,~\cite{Xu2011a}
which are unlike those in the random field or the random bond disorder models with short-range interactions.~\cite{Imry1975a, Blinc1999a, Westphal1992a} 

Typically only such interaction models with quenched disorder have been used to describe relaxor ferroelectrics. 
These models in three dimensions for weak disorder also have long-range order 
at a finite temperature,~\cite{Imry1975a} which is not observed in the canonical relaxor PMN.~\cite{Cowley2011a}
Several other models have been proposed for relaxors,~\cite{Glinchuk2004a, Burton2006a, Tinte2006a, Akbarzadeh2012a, Takenaka2013a, Sherrington2014a}
 - including those of Burton et al.~\cite{Burton2006a} and of Tinte et al.~\cite{Tinte2006a} which
just like it is done in the present paper, considers well-known effective Hamiltonians for 
conventional ferroelectrics with quenched disorder - 
however, they have almost exclusively been solved by numerical methods, which 
makes it difficult to determine general aspects of the solution of such models.

Recently, it has been proposed that the ground state of relaxors is that of a cluster glass in which polar 
domains generated by random fields cluster at low temperatures due to frustrating nature of the dipolar force.~\cite{Kleemann2014a}
The mechanism by which such a glassy state is formed is described by the subsequent formation of polar domains due to random electric 
fields and the clustering of mesoscopic domains due to frustrated random dipolar interactions. 
While we do not study this problem here, we cannot rule out that such freezing 
mechanisms may be out of reach of  the theoretical treatment that is presented here, as 
nucleation of polar domains within the disordered matrix may occur as there are stable ordered states in the 
free energy of our model.~\cite{Guzman2013a} Such domains would then interact through the dipolar force, 
which is in itself frustrating due to its anisotropy.

In a previous paper,~\cite{Guzman2013a}
we developed a variational method which allowed us to 
study the temperature evolution of the free energy 
with compositional disorder, which is essential to understand 
the dielectric properties of solid solutions of relaxors with conventional ferroelectrics such 
as PbMg$_{1/3}$Nb$_{2/3}$O$_3$-PbTiO$_3$~(PMN-PT). We found that there are disordered states
with a region of metastability that extends to zero temperature 
for moderate disorder and that  field-induced transitions
to stable ferroelectric states can occur only for applied electric fields sufficiently large 
to overcome the energy barriers that result from disorder. 
Here, we  present a different solution of the model based on Onsager's results on systems with dipolar interactions~\cite{Onsager1936a} 
that provides an important physical insight that is not obvious from the variational solution.

In formulating the minimum necessary model for PMN, it is important to recall Onsager's result~\cite{Onsager1936a} 
that unlike in the Clausius-Mossotti or Lorentz approximation, dipole interactions alone do not lead to ferroelectic 
order except at $T=0$. Moreover, pure ferroelectric transitions were understood with the realization~\cite{Cochran-Anderson}  
that they are soft transverse optic mode transitions 
due to dipoles induced by structural transitions 
so the low temperature phase 
does not have a center of symmetry. 
The first point is not in practice important for pure ferroelectrics which are well 
described by a mean-field theory for the dielectric constant, and is often a first order transition 
only below which the dipoles are produced.~\cite{Lines1977a} But as 
we show below, in the relaxor PMN the random location of defects  acts in concert 
with the dipole interactions to extend the region of fluctuations to zero temperatures. Therefore dipole 
interactions must be a necessary part of the model. 
The essential physical points in the simplest necessary solution is to formulate a theory which considers thermal 
and quantum fluctuations at least at the level of the Onsager approximation~\cite{Onsager1936a} and random field 
fluctuations at least at the level of a replica theory.~\cite{DeDominicis_book} 

\section{Model Hamiltonian}
We consider the model Hamiltonian presented in Ref.~[\onlinecite{Guzman2013a}], which we present
here for the sake of completeness. 
We focus on the relevant transverse optic mode configuration coordinate  $u_i$ of the ions in the unit cell $i$ along 
the polar axis (chosen to be the $z$-axis).  $u_i$ experiences a local random field $h_i$ with probability $P(\{h_{i}\})$ 
due to the compositional disorder introduced by the different ionic radii and different valencies of Mg$^{+2}$, Nb$^{+5}$ in PMN. The model Hamiltonian is 
\begin{align}
\label{eq:Hamiltonian_RFE}
H&=\sum_{i}\left[\frac{\Pi_i^2}{2M}+V(u_i)\right]-\frac{1}{2}\sum_{i,j}v_{ij}u_i u_j  \nonumber \\
 &~~~~~~~~~~~~~~~~~~~~~~~~~~~~~~~~
 -\sum_i E_i^{ext} u_i-\sum_i h_i u_{i},
\end{align} 
where $\Pi_i$ is the momentum conjugate to $u_i$, $M$ is an effective mass and we have included an external electric field 
$E_i^{ext} = E_0^{ext} + E_i^{ext}(t)$, with a static and uniform component $E_0^{ext}$ and 
an infinitesimally small time-dependent component, $E_i^{ext}(t)$. 
We assume the $h_i$'s are independent random variables with zero mean and variance $\Delta^2$.
$V(u_i)$ is an anharmonic potential,
\begin{align}
\label{eq:anharmonic_potential_RFE}
V(u_i)=\frac{\kappa}{2} u_i^2 + \frac{\gamma}{4} u_i^4,
\end{align}
where $\kappa,~\gamma$ are positive constants.
$v_{ij}$ is the dipole interaction,
\begin{equation}
\label{eq:vij}
v_{ij}/{e^*}^2=
\begin{cases}
3 \frac{(Z_{i}-Z_{j})^2}{|{\bm R}_i-{\bm R}_j|^5} -\frac{1}{|{\bm R}_i-{\bm R}_j|^3}, & {\bm R}_i \neq {\bm R}_j\\
0, & {\bm R}_i = {\bm R}_j,
\end{cases}
\end{equation}
where $e^*$ is the Born effective charge and $Z_i$ is the $z$-component of ${\bm R}_i$.
The Fourier transform of the dipole interaction 
$ v_{\bm q} / ( n {e^*}^2 ) = 1/ ( n {e^*}^2 )  \sum_{i,j}v_{ij} e^{ i {\bm q} \cdot ( {\bm R}_i - {\bm R}_j ) }= \frac{4\pi}{3}\left(1-3 \frac{q_z^2}{|{\bm q}|^2}\right)-\zeta |{\bm q} a|^2 + 3\zeta (q_z a)^2 $ 
is non-analytic for ${\bm q} \to 0$.
For future use, we denote $v_{0}^{\perp} = 4\pi n {e^*}^2 / 3 $ 
as the ${\bm q} \to 0 $ component of $v_{\bm q}$ in the direction transverse to the polar axis~(${\bm q} \perp \hat{ \bm z })$,  
$n$ as the number of unit cells per unit volume, and $a$ as the lattice constant. 
$\zeta$ is a dimensionless coefficient that depends on the structure of the lattice.~\cite{Aharony1973a}

The Hamiltonian of Eq.~(\ref{eq:Hamiltonian_RFE}) presents 
(i) long ranged (anisotropic) dipolar interactions; (ii) compositional disorder; and (iii) anharmonicity. We now present the 
procedure used to find the correlation functions for (\ref{eq:Hamiltonian_RFE}). 
We use the quasi-harmonic approximation, similar to that used for pure displacive ferroelectrics to treat anharmonicity.~\cite{Lines1977a}
 This reduces the problem to that of harmonic oscillators with a renormalized stiffness which is determined self-consistently. Disorder is treated by
an approach which may be termed the ``poor man's replica method", and has been used to study two and three dimensional magnets 
with quenched random fields.~\cite{Vilfan1985a} The Onsager approximation, which may be considered the leading fluctuation 
correction over the mean-field approximation~\cite{VanVleck1937a} is used for the dipolar interactions as well as for the general statistical-mechanical treatment of the problem. 

\section{Solution by Onsager Cavity Method}

In this approximation,  
\begin{align}
\label{eq:Onsager_Hamiltonian_RFE}
H & \approx & \sum_i H_i = \sum_i \left( H_i^0 - h_i u_i - E_i^{cavity} u_i \right),
\end{align}
with $H_i^0 = \Pi_i^2/(2M) + V(u_i) - (\lambda/2) u_i^2$. $E_i^{cavity}$ is the Onsager cavity field,
\begin{equation}
\label{eq:cavity}
E_i^{cavity}  = \sum_{j}v_{ij}\bra u_j \ket - \lambda \bra u_i \ket +  E_i^{ext}.
\end{equation}
\noindent
$ \bra X \ket$ denotes the thermal average of any quantity $X$; similarly we will denote by $ \ovl{ \bra X \ket } $, 
the configurational average over the quenched random fields $h_i$ taken {\it after} the thermal average is taken.

Onsager calculated the parameter $\lambda$ for the simpler problem of dipolar forces alone using continuum electrostatics. 
More generally, one may determine
$\lambda$ by use of a self-consistent procedure such that the response functions obey the fluctuation dissipation theorem.~\cite{Brout1967a} 
To find the self-consistent response functions, we consider the thermal average of $u_i$ made up
of a static and uniform order-parameter $p = \ovl{\bra u_i \ket} $ and a linear response part $ \delta \bra  u_i \ket $, i.e.,  $ \bra u_i \ket = p + \delta \bra u_i \ket$.
The linear response is determined from the cavity field $E_i^{cavity}$,~\cite{Lines1977a}
\begin{align}
\delta \bra  u_i \ket  = \phi_{h_i}(\omega) \delta E_i^{cavity},
\end{align}
where $\phi_{h_i}(\omega)$ is the dielectric susceptibility for the problem with Hamiltonian, $H_i^0 - u_i h_i - u_i p (v_0^\perp - \lambda) - u_i E_0$
and $ \delta E_i^{cavity}  = \sum_j v_{ij} \delta \bra  u_j \ket - \lambda \delta \bra  u_i \ket + \delta E_i^{ext}(t)$.
To take the average over configurations, we assume that the effects of the random fields $h_i$ decouple in $\phi_{h_i}(\omega)$ 
and $ \bra u_i \ket$. For our model, as will be shown below, $ \ovl{ \phi_{h_i} }(\omega) $ is  independent of $i$ and 
therefore has only a ${\bm q}=0$ component denoted simply by $ \ovl{ \phi_{h}}(\omega)  $. Then taking a Fourier transform, 
we get the formal expression for the susceptibility function,
\begin{equation}
\label{eq:dynamic_susceptibility_RFE_1}
\chi_{\bm q}(\omega) =  \frac{ \ovl{ \delta \bra  u_{\bm q} \ket }  }{ \delta E_{\bm q}^{ext}} = \frac{\ovl{ \phi_{h}}(\omega)}{ 1  + \ovl{ \phi_{h}}(\omega) \left( v_{\bm q} - \lambda \right)} 
\end{equation}
Next we determine $\phi_{h}(\omega)$. The equation of motion for $u_i$ due to $H_0 - h_i u_i - u_i p (v_0^\perp - \lambda) - u_i E_0^{ext} $ 
and including a damping force 
characterized by $\Gamma$ are
\begin{align}
\label{eq:eq_of_motion_non_interacting_RFE_1}
M \ddot{u}_i &= -\frac{d V(u_i)}{du_i}  +  \lambda u_i + h_i + p (v_0^\perp - \lambda)  \nonumber \\ 
&~~~~~~~~~~~~~~~~~~~~~~~~~~~~~~~~~~~~~~~~~~
+  E_0^{ext} - \Gamma \dot{u}_i. 
\end{align}
We now linearize (\ref{eq:eq_of_motion_non_interacting_RFE_1})
by considering fluctuations $\delta u_i$ around the static and uniform expectation value of $u_i$, i.e.,  
$u_i = p + \delta u_i$ to get that 
\begin{align}
\label{eq:phi_RFE_1}
 \ovl{ \phi_{h} }(\omega)  &=  \frac{1}{ {M\Omega^\prime}^2 - \omega^2 + i \omega \Gamma }, \\
\label{eq:phi_RFE_2}
 E_0 &= \left[ M {\Omega^\prime}^2  - ( v_0^\perp - \lambda  ) - 2\gamma p^2 \right] p,
\end{align}
where $M {\Omega^\prime}^2  \equiv \kappa -\lambda + 3 \gamma \left( \ovl{ \bra (\delta u_i)^2  \ket } + p^2 \right)$.

The fluctuations $ \ovl{ \bra (\delta u_i)^2  \ket } $ of the $H_0-h_iu_i$ problem are easily shown to be,
\begin{eqnarray}
\label{eq:eta_RFE_1}
\ovl{ \bra (\delta u_i)^2  \ket }  &=&  \frac{ \hbar }{2 M \Omega^\prime} \coth\left( \frac{ \beta \hbar \Omega^\prime }{2} \right) + \ovl{ \bra u_i \ket^2 } - p^2. 
\end{eqnarray}

We finally find the susceptibility~(\ref{eq:dynamic_susceptibility_RFE_1}),
\begin{equation}
\label{eq:susceptibility_RFE_2}
\chi_{\bm q}(\omega) =  \frac{1}{ M \Omega_{\bm q}^2 - M\omega^2 + i \omega \Gamma }, 
\end{equation}
where $ \Omega_{\bm q} $ is the vibration frequency of $u_i$ for the full problem,
\begin{equation}
\label{eq:omega_q_RFE_1}
M \Omega_{\bm q}^2  = M \left( \Omega_0^\perp \right)^2 + \left( v_0^\perp - v_{\bm q} \right),
\end{equation}
and $\Omega_0^\perp$ is the ${\bm q} = 0$ component of the phonon frequency in the direction 
perpendicular to the polar axis,
\begin{align}
\label{eq:omega_0_RFE_1}
M \left( \Omega_0^\perp \right)^2  = M {\Omega^\prime}^2 - \left( v_0^\perp - \lambda \right).
\end{align}
\noindent
We now determine 
the parameter $\lambda$ by enforcing the fluctuation dissipation theorem,
\begin{eqnarray}
\label{eq:fluctuation_dissipation_RFE_1}
\ovl{ \bra u_i^2 \ket}   && - \ovl{ {\bra u_i \ket }^2 } =  \nonumber \\
&& \hspace{-0.75cm}\frac{1}{N}\sum_{\bm q} \frac{1}{2\pi} \int_{-\infty}^{\infty} d\omega \coth \left( \frac{\beta \hbar \omega}{2} \right) \mbox{Im}\left[ \chi_{\bm q}(\omega)  \right]. 
\end{eqnarray}
The summation over ${\bm q}$ extends over the first Brillouin zone. 
To close the system of equations the polarization $\bra u_i \ket_T^0$ must itself be related to
the random fields through the susceptibility~(\ref{eq:susceptibility_RFE_2}). For a fixed
 realization of disorder,~\cite{Vilfan1985a}
\begin{equation}
\label{eq:avg_1_RFE_1}
\bra u_i \ket = p + \sum_{j} \ovl{\chi}_{ij} h_j,
\end{equation}
where $ \ovl{ \chi}_{ij} (0)$ is the zero-frequency susceptibility averaged over compositional disorder
with the Fourier transform $ \chi_{\bm q} (0)= \sum_{ij} \ovl{ \chi}_{ij}(0) e^{ i {\bm q} \cdot ( {\bm R}_i - {\bm R}_j ) } $.
\noindent
Using $ \ovl{ h_j }  =  0 $ and $\ovl{ h_i h_j } = \Delta^2 \delta_{ij}$,
\begin{equation}
\label{eq:avg_2_RFE_1}
\ovl{ \bra u_i  \ket } = p, ~~
\ovl{ \bra u_i  \ket^2 } = p^2 + \frac{\Delta^2}{N}\sum_{\bm q} \chi_{\bm q}^2(0).
\end{equation}

Given the above solution, the temperature and disorder dependence of the 
the dielectric susceptibility $\chi_{\bm q}(\omega)$ given by Eq.~(\ref{eq:susceptibility_RFE_2}),
the phonon frequency $\Omega_{\bm q}$ of Eq.~(\ref{eq:omega_q_RFE_1}),
and the static polarization $p$ of Eq.~(\ref{eq:avg_2_RFE_1}) can be determined self-consistently
together with the parameter $\lambda$ in Eq.~(\ref{eq:fluctuation_dissipation_RFE_1}). 
It is easy to show that by eliminating $\Omega^\prime$ and $\lambda$ from Eqs.~(\ref{eq:phi_RFE_2}),~(\ref{eq:eta_RFE_1}), and (\ref{eq:omega_0_RFE_1})
one recovers the results of Ref.~[\onlinecite{Guzman2013a}].

The static structure factor $S_{\bm q}$ is derivable from $\chi_{\bm q}(\omega)$ by use of the standard procedure.~\cite{Halperin1976a} 
We obtain the following result,
\begin{equation}
\label{eq:static_Sq_RFE_1}
S_{\bm q} 
= p^2 \delta_{\bm q} +
\frac{\hbar}{2 M \Omega_{\bm q}} \coth \left( \frac{\beta \hbar \Omega_{\bm q}}{2} \right) + \frac{\Delta^2}{  \left( M \Omega^2_{\bm q}\right)^2 },
\end{equation}
where the transverse optic phonon frequency $\Omega_{\bm q}$ is given in Eq.~(\ref{eq:omega_q_RFE_1}) and it is calculated 
self-consistently as described above. In the absence of disorder and in the classical limit~($\hbar \to 0$),
we recover the standard results for conventional ferroelectrics.~\cite{Lines1977a} 
The line shape of the structure factor~(\ref{eq:omega_q_RFE_1}) 
resembles that of the well-known Lorentzian plus Lorentzian squared for disordered magenetic systems~\cite{Belanger1992a} 
with the important distinction that the relevant interactions in our problem are long-ranged and anisotropic
dipolar forces rather than short-ranged isotropic exchange interactions.

\section{Results}

\begin{figure}[htp]
\begin{centering}
\includegraphics[scale=0.4]{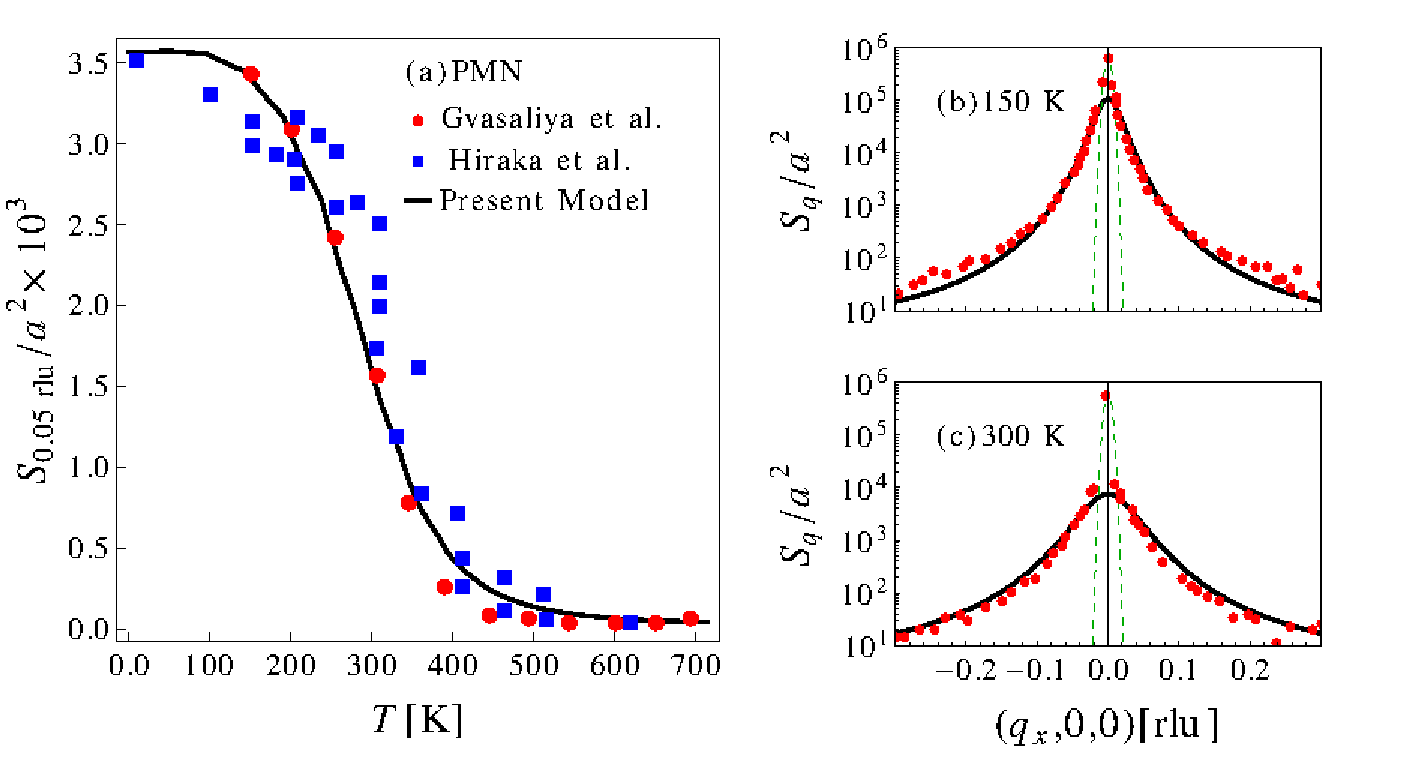}
\caption{(a) Temperature and (b)-(c) wavevector dependence of the structure factor $S_{\bm q}$.
Red dots and blue squares correspond to neutron scattering data from Refs.~[\onlinecite{Gvasaliya2005a, Hiraka2004a}]
in the vicinity of the $(110)$ and $(100)$ Bragg peaks in PMN.
Black solid line corresponds to the calculated $S_{\bm q}$
with $(v_0^\perp-\kappa)/{v_0^\perp}=0.85,~\gamma a^2/v_0^\perp = 0.09,~\hbar / ( M v_0^\perp a^4 )^{1/2} =5.0,~\Delta^2/(v_0^\perp a)^2 = 0.03$,
and $k_B T_c/ (v_0^\perp a^2 ) = 1.0$ with $T_c = 465\,$K.
The green dashed line is a Gaussian Bragg peak of width given by the experimental resolution used in Ref.~[\onlinecite{Gvasaliya2005a}]~($\simeq 0.01\,$rlu).}
\label{fig:S0_vs_T}
\end{centering}
\end{figure}

We now present results of the calculations to illustrate the physical principles. 
We focus on the disordered states~($p=0$) as the experiments we will compare have been performed in such a phase.
The model parameters are obtained from fits to the experimental structure factor of PMN~(see Fig.~\ref{fig:S0_vs_T}).

\begin{figure}[htp]
\begin{centering}
\includegraphics[scale=0.75]{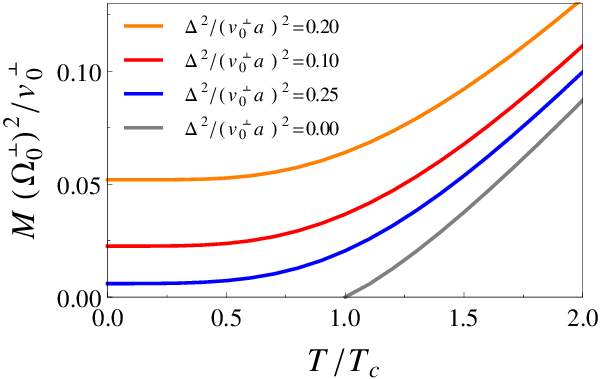}
\caption{Temperature dependence of the zone center optic phonon frequencies $\Omega_0^\perp$ for 
dipole interactions and compositional disorder. The frequencies remain finite down to $T=0$ for finite disorder, thus
excluding long-range ferroelectric order.}
\label{fig:omega_RFE}
\end{centering}
\end{figure}

Figure~\ref{fig:omega_RFE} presents the calculated temperature dependence of the (mean) transverse optical model frequency  
at ${\bm q=0}$ suitably normalized to show its different regimes. The frequency remains finite at all temperatures 
even for weak disorder, thus excluding long-range order in the model which includes dipolar interactions and weak disorder~(long-range 
order would be present for the model with short-range interactions alone).~\cite{Imry1975a} 
This state is metastable at low temperatures.~\cite{Guzman2013a} 
As expected, these results are similar to those found previously~\cite{Guzman2013a} and 
are shown here for the sake of completeness.

\begin{figure}[htp]
\begin{centering}
\includegraphics[scale=0.75]{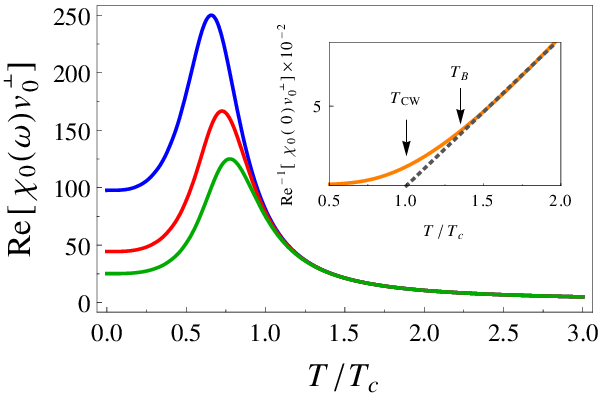}
\caption{Temperature and frequency dependence of the real part of the dynamic susceptibility $\chi_{0}(\omega)$.
Inset: Inverse of the real part of the static susceptibility
with compositional disorder~($\Delta^2/(v_0^\perp a)^2 = 0.03$). Deviations from Curie-Weiss law are indicated by the gray dashed line.
Blue~(top), red~(middle), and green~(bottom) lines correspond to $\omega \Gamma /v_0^\perp = 2.0\times 10^{-2},~3.0\times 10^{-2},~4.0\times 10^{-2}$.}
\label{fig:dielectric_constant_RFE}
\end{centering}
\end{figure}

\begin{figure}[htp]
\begin{centering}
\includegraphics[scale=0.75]{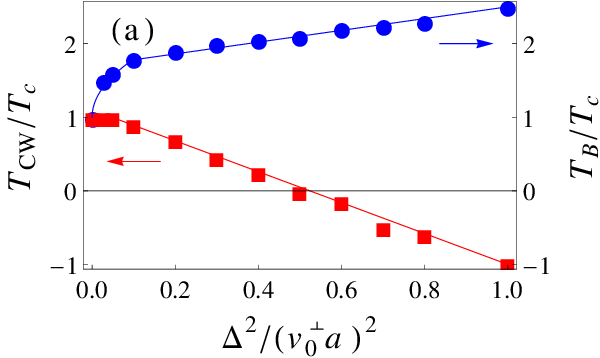}
\includegraphics[scale=0.75]{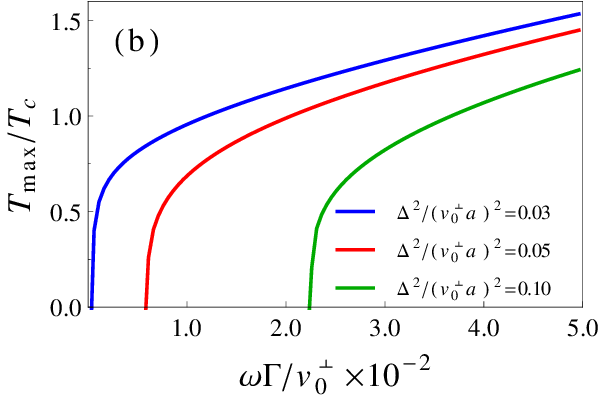}
\caption{(a) Curie-Weiss~($T_{CW}$) and Burns~($T_B$) temperatures dependence on compositional disorder.
(b) Frequency and disorder dependence of the temperature $T_{\max}$ at which the dielectric constant is maximum. }
\label{fig:Tcw_T_B}
\end{centering}
\end{figure}

\begin{figure*}[htp]
\begin{center}
\includegraphics[scale=0.5]{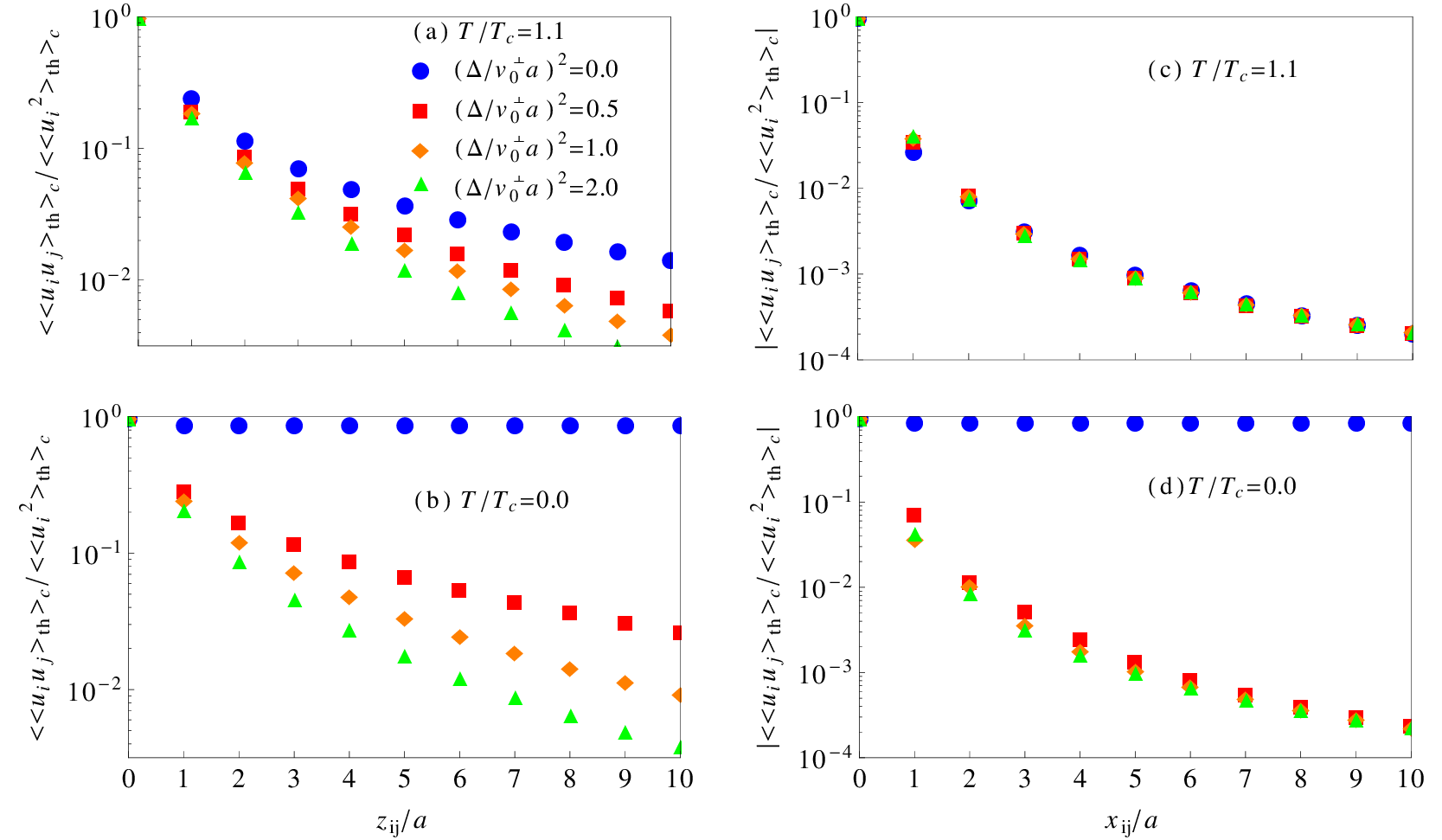}
\caption{
Log-linear plot of near-neighbor correlation functions of polarization for several temperatures
and disorder strengths. Here, $(v_0^\perp-\kappa)/v_0^\perp=0.40,~\gamma a^2/v_0^\perp = 0.10$, and $ \hbar / ( M v_0^\perp a^4 )^{1/2} =1.0$.}	
\label{fig:nn_correlations_w_disorder}
\end{center}
\end{figure*}

Associated with the phonon frequency is the 
dielectric response $\chi_0(\omega)$, shown in Fig.~\ref{fig:dielectric_constant_RFE}, where $T_{CW}$  and $T_B$  are also 
identified (see the inset in Fig.~\ref{fig:dielectric_constant_RFE}). 
Figs.~\ref{fig:Tcw_T_B}~(a)-(b) give the variation of $T_{CW}$, $T_B$, and $T_{\max}$ with the parameters of the model. 
We see that the characteristic temperatures are understood by the ratio of the disorder distribution to the transverse
dipole interaction with the damping $\Gamma$ playing a major role for $T_{\max}$. Clearly,
our model is too simple to give the dynamics observed in the relaxor susceptibility  
such as the Vogel-Fulcher behavior.~\cite{Glazounov1998a} Nonetheless, it can capture several of the temperature scales. 
To get detailed dynamics, one must add the full landscape of potentials and relaxation processes, which
have already been considered in the literature.~\cite{Burton2006a, Takenaka2013a} This is not the aim of the present paper
which is concerned  with the simpler question of static structure factor. We may add, however,
that it is necessary to have the theory of static structure factor well in hand before one can
reliably consider the more complicated dynamical problems. 

We now compare the static structure factor  $S_{\bm q}$  with that measured by neutron scattering experiments~\cite{Gvasaliya2005a, Hiraka2004a}.
Without compositional disorder~($\Delta=0$), simple inspection shows that
$S_{\bm q=0}$ diverges as the mode frequency softens in the vicinity 
of the critical point $T_c$ where long-range ferroelectric order sets in.
For finite compositional disorder~($\Delta > 0$), 
$S_{\bm q=0}$ remains finite at all temperatures and there is no long-range ferroelectric order.
We find this temperature behavior is in agreement with that observed by neutrons in PMN, as shown in 
Fig.~\ref{fig:S0_vs_T}~(a).
The flat behavior of $S_{\bm q}$ at low temperatures~($T \to 0 $) is due to
zero-point fluctuations: in the classical limit~($\hbar \to 0$), $S_{\bm q}$ of Eq.~(\ref{eq:static_Sq_RFE_1}) increases with decreasing temperature and is finite at $T=0$. 

Figures~\ref{fig:S0_vs_T}~(b)-(c) compare the calculated wavevector distribution 
of $S_{\bm q}$ with that  
observed for the relaxor PMN at $T=150\,$K and $T=300\,$K. We use the same value of the model parameters as those of Fig.~\ref{fig:S0_vs_T}~(a).
The observed line shape cannot be described by a simple Lorentzian seen for conventional perovskite ferroelectrics or the 
``squared Lorentzian" expected for random field models.

We now compute the spatial dependence and anisotropy of the correlation functions of polarization at
various characteristic temperatures and various normalized disorder strengths. These spatial correlations 
have been Fourier transformed to give the static structure factor $S_{\bm q}$.
One of the purposes of this section is to contrast the correlations of our model to those expected from hypothetical
polar nanoregions~(PNRs) on which we will comment at the end. 

We first consider the correlation functions of polarization at large distances~($|{\bm R}_{ij}| \to \infty$). 
For low temperatures~($T \to 0$) and arbitrarily small but finite disorder~($0 < \Delta/(v_0^\perp a) \ll 1$), 
the correlations are anisotropic and slowly decaying functions,
\begin{equation}
\label{eq:correlations_RFE_3}
\ovl{ \bra u_i u_j \ket } = 
\begin{cases}
\left(\frac{4\pi}{3}\right)^2 \frac{ \Delta^2 }{\left( v_0^\perp \right)^2} \times \frac{ 32\pi^3 }{ v_{BZ} } \frac{ (\bar{\chi}_0^\perp)^{3} }{ |{\bm R}_{ij}|^3 }, & {\bm R}_{ij} \parallel \hat{\bm z}, \\
-\left(\frac{4\pi}{3}\right)^2 \frac{ \Delta^2 }{\left( v_0^\perp \right)^2} \times \frac{ 1 }{2} \frac{ \pi^{3/2} }{ v_{BZ} } \frac{ (\bar{\chi}_0^\perp)^{3/2} }{ |{\bm R}_{ij}|^3 }, &  {\bm R}_{ij} \perp \hat{\bm z},
\end{cases}
\end{equation}
where $ \left( \bar{\chi}_0^\perp \right)^{-1} = (4\pi/3) \left( \chi_0^\perp \right)^{-1} / v_0^\perp $ is a small but finite dimensionless inverse susceptibility; and
$v_{BZ}$ is the volume of the Brillouin zone.
The ratio of the longitudinal to transverse components is proportional (in absolute value) to $ \left(\bar{\chi}_0^\perp\right)^{3/2} $, indicating
that the positive longitudinal correlations are stronger than the negative transverse components.
This behavior is similar to that of ferroelectrics without random fields. 
For an uniaxial ferroelectric without compositional disorder and above the critical temperature $T_c$, 
the correlation functions exhibit the same anisotropy, power law decay, and 
longitudinal to transverse ratio~\cite{Lines1974a}. 
The correlations of Eq.~(\ref{eq:correlations_RFE_3}) are, however, 
in sharp contrast with those of short-range interactions with quenched random fields in 
three dimensions where the correlation functions decay exponentially.~\cite{Birgeneau1983a}

We now discuss the near-neighbor correlations with compositional disorder.
For temperatures slightly above $T_c$, 
the near-neighbor longitudinal correlations with and without disorder show similar
decay with an overall strength that decreases with increasing compositional disorder~(Fig.~\ref{fig:nn_correlations_w_disorder}(a)).
This behavior persists down to $T=0$ for finite compositional disorder~(Fig.~\ref{fig:nn_correlations_w_disorder}(b)). 
For no disorder, the correlations exhibit the long-range order expected at $T=0$.
The correlations in the transverse direction 
follow similar decay as that of the longitudinal components except 
that they are negative and significantly weaker~(Figs.~\ref{fig:nn_correlations_w_disorder}(c)-(d)).

To compare with the idea of PNRs invoked to rationalize the behavior of relaxors, we note that we find
that significant correlations develop below the Burns temperature $T_B$. However, we do not find a difference in the characteristic
behavior at short distances compared to that at long distances, which might have been expected for PNRs.
We find  that the correlations have a power law behavior which joins smoothly to a short-range part where they must saturate
to near the on-site correlations. This is characteristic of the fluctuation regime of any cooperative problem. We have emphasized that relaxors may be
looked on as materials in which the fluctuation regime extends from the Burns temperature all the way to $T=0$.
We therefore conclude that the diffuse scattering observed by neutrons does not support the qualitative picture of PNRs. 
Recently, other works have arrived at similar conclusions.~\cite{Bosak2012a, Hlinka2012a, Takenaka2013a}
We point out, however, that nucleation of local polar domains within the non-polar
phase may occur as there are stable ferroelectric states
in the free energy.~\cite{Guzman2013a}

\section{Conclusions}

We have used the insight that the frustrating nature of the dipolar interactions (due to their anisotropy) introduced in a perovskite
due to a putative displacive transition together with quenched disorder impedes a ferroelectric transition at all temperatures and 
leads instead to a region of extended ferroelectric fluctuations 
in PMN. The nature of the dipolar interactions is such that the physics cannot be captured in a 
mean-field like or two body correlation approximations to the problem. 
Using a minimal approximation scheme which is able to handle the 
special nature of the problem, we are able to derive the observed structure factor of 
the relaxor PMN and relate it to their static and dynamic microscopic properties.

In addition to the difficulties posed to our current theoretical treatment 
by the complex dynamic processes of relaxors (see Sec. IV),  there are several other challenges
which should be addressed in future extensions of this work such as the effects of cubic symmetry, 
of disorder in the bonds (e.g. lattice stiffness and dipole interactions), and of electrostriction. 
These are all important ingredients of any model that aims to describe of the universality class,~\cite{Cowley2011a} 
glassiness,~\cite{Kleemann2014a} and ultra-high piezoelectricity~\cite{piezoelectricity} of typical relaxors such as 
PMN and its solid solutions with conventional ferroelectrics such as PMN-PT.

\section{Acknowledgments}
This research was partially supported by the University of California Lab Fee Program 09-LR-01-118286-HELF.
We thank Peter Littlewood for insightful discussions and other principal investigators with whom this 
grant was issued: Frances Hellman, Albert Migliori and Alexandra Navrotsky.
Work at Argonne National Laboratory is supported by the U.S. Department of Energy, 
Office of Basic Energy Sciences under contract no. DE-AC02-06CH11357, 
and work at the University of Costa Rica by Vicerrector\'{i}a de Investigaci\'{o}n
under the project no. 816-B5-220.


\begin{thebibliography}{99}

\bibitem{Cowley2011a}
R. A. Cowley, S. N. Gvasaliya, S. G. Lushnikov, B. Roessli and G. M. Rotaru, Adv. Phys. \textbf{60}, 229 (2011).

\bibitem{Bokov2006a}
A. A. Bokov and Z.-G. Ye, J. Mater. Sci. \textbf{41}, 31 (2006).

\bibitem{Kleeman2006a}
W. Kleeman, J. Mater. Sci. \textbf{41}, 129 (2006).

\bibitem{Samara2003a}
G. A. Samara, J. Phys.: Condens. Matter \textbf{15}, R367 (2003).

\bibitem{Cross1987a}
L. E. Cross, Ferroelectrics \textbf{151}, 305 (1994);
L. E. Cross, Ferroelectrics \textbf{76}, 241 (1987).

\bibitem{Burns1983}
G. Burns and B. A. Scott, Solid State Commun. \textbf{13}, 423 (1973);
G. Burns and F. H. Dacol, Solid State Commun. \textbf{48}, 853 ͑(1983͒);
G. Burns and F. H. Dacol, Phys. Rev. B \textbf{28}, 2527 ͑(1983͒).

\bibitem{Viehland1992a}
D. Viehland, S. J. Jang, L. E. Cross and M. Wuttig, Phys. Rev. B \textbf{46} 8003 (1992).

\bibitem{Smolenskii1970a}
G. A. Smolenskii, J. Phys. Soc. Jpn. \textbf{28}, Suppl. 26 (1970).

\bibitem{Bovtun2004a}
V.~Bovtun, S.~Kamba, A.~Pashkin, M.~Savinov, P.~Samoukhina, J.~Petzelt,
I.~P.~Bykov and M.~D.~Glinchuk, Ferroelectrics \textbf{298}, 23 (2004).

\bibitem{Hlinka2006a}
J. Hlinka, J. Petzelt, S. Kamba, D. Noujni and T. Ostapchuk, Phase Transitions \textbf{79}, 41 (2006).

\bibitem{Bokov2012a}
A. A. Bokov and Z.-G. Ye J. Adv. Dielec. \textbf{2}, 1241010 (2012).

\bibitem{Glazounov1998a}
A.E. Glazounov and A.K. Tagantsev, Appl. Phys. Lett. \textbf{73}, 856 (1998).

\bibitem{Gvasaliya2005a}
 S. N. Gvasaliya, B. Roessli, R. A. Cowley, P. Huber and S. G. Lushnikov, 
J. Phys.: Condens. Matter \textbf{17}, 4343 (2005);
R.~A.~Cowley,~S. N.~Gvasaliya and B.~Roessli, 
Ferroelectrics \textbf{378}, 53 (2009).

\bibitem{Hiraka2004a}
H. Hiraka, S.~-H Lee, P.M.~Gehring, Guangyong~Xu and G.~Shirane, Phys. Rev. B \textbf{70}, 184105 (2004).

\bibitem{Gehring2009a}
P. M. Gehring, H. Hiraka, C. Stock, S.-H. Lee, W. Chen, Z.-G. Ye, S. B. Vakhrushev, and Z. Chowdhuri, 
Phys. Rev. B \textbf{79} 224109 (2009).

\bibitem{Xu2011a}
for reviews, see G. Xu, J. Phys. Conf. Ser. \textbf{320}, 012081 (2011) 
and P. M. Gehring, J. Adv. Dielec. \textbf{2}, 1241005 (2012).

\bibitem{Imry1975a}
Y. Imry, S. Ma, Phys. Rev. Lett. \textbf{35}, 1399 (1975).

\bibitem{Blinc1999a}
R. Pirc, R. Blinc, Phys. Rev. B \textbf{60}, 13470 (1999);
R. Blinc, J. Dolinsek, A. Gregorovic, B. Zalar, C. Filipic, Z. Kutnjak, A. Levstik, and R. Pirc,
Phys. Rev. Lett. \textbf{83},  424 (1999).

\bibitem{Westphal1992a}
V. Westphal, W. Kleemann, M. D. Glinchuk, Phys. Rev. Lett. \textbf{68}, 847 (1992).

\bibitem{Glinchuk2004a}
M. D. Glinchuk, Br. Ceram. Trans. \textbf{103}, 76 (2004).

\bibitem{Burton2006a}
B. P. Burton, E. Cockayne, S. Tinte, and U. V. Waghmare, Phase Transitions \textbf{79}, 91 (2006).

\bibitem{Tinte2006a}
S. Tinte, B. P. Burton, E. Cockayne, and U. V. Waghmare Phys. Rev. Lett. \textbf{97}, 137601 (2006).

\bibitem{Akbarzadeh2012a}
A. R. Akbarzadeh, S. Prosandeev, E. J. Walter, A. Al-Barakaty, and L. Bellaiche, Phys. Rev. Lett. \textbf{108}, 257601 (2012).

\bibitem{Takenaka2013a}
H. Takenaka, I. Grinberg, and A. M. Rappe, Phys. Rev. Lett. \textbf{110}, 147602 (2013). 

\bibitem{Sherrington2014a}
D. Sherrington, Phys. Rev. B \textbf{89}, 064105 (2014).

\bibitem{Kleemann2014a}
W. Kleemann, in Mesoscopic Phenomena in Multifunctional Materials, edited by A. Saxena and A. Planes, Springer Series in Materials Science, Vol 198, p. 249 (Springer, Berlin, 2014).

\bibitem{Guzman2013a}
G.~G.~Guzm\'{a}n-Verri, P. B. Littlewood, and C.~M.~Varma, Phys. Rev. B \textbf{88}, 134106 (2013). 

\bibitem{Onsager1936a}
L. Onsager, J. Am. Chem. Soc. \textbf{58}, 1486 (1936).

\bibitem{Cochran-Anderson}
W. Cochran, Phys. Rev. Lett. \textbf{3}, 412 (1959); W. Cochran, Adv. Phys. \textbf{9}, 387 (1960);
P. W. Anderson,	Fizika Dielectrikov, ed. G. I. Skanavi (Akad. Nauk SSSR Fizicheskii Inst., im P. N. Levedeva, Moscow, 1960);
P. W. Anderson, {\it A Career in Theoretical Physics}, (World Scientific Publishing Co., New Jersey, 1994).


\bibitem{Lines1977a}
M. E. Lines and A. M. Glass, {\it Principles and Applications of Ferroelectrics and Related Materials} 
(Clarendon Press, Oxford, 1977).

\bibitem{DeDominicis_book}
C. De Dominicis and I. Giardina, {\it Random Fields and Spin Glasses: A Field Theory Approach}~(Cambridge University Press, Cambridge, 2010).

\bibitem{Aharony1973a}
A. Aharony and  M. E. Fisher, Phys. Rev. B \textbf{8}, 3323 (1973).

\bibitem{Vilfan1985a}
I. Vilfan and R. A. Cowley, J. Phys. C: Solid State Phys. \textbf{18}, 5055  (1985).

\bibitem{VanVleck1937a}
J. H. Van Vleck, J. Chem. Phys.~\textbf{5}~320~(1937).

\bibitem{Brout1967a}
R. Brout and H. Thomas, Phys. \textbf{3}, 317 (1967); H. Thomas and R. Brout, J. Appl. Phys. \textbf{39}, 624 (1968).

\bibitem{Halperin1976a}
B. I. Halperin and C. M. Varma, Phys. Rev. B \textbf{14}, 4030 (1976).

\bibitem{Belanger1992a}
D. P. Belanger and A. P. Young, J. Magn. Magn. Mater. \textbf{100}, 272 (1992).

\bibitem{Lines1974a}
M. Lines, Phys. Rev. B \textbf{9}, 950 (1974).

\bibitem{Birgeneau1983a}
R. J. Birgeneau, H. Yoshizawa, R. A. Cowley, G. Shirane and H. Ikeda, Phys. Rev. B \textbf{28}, 1438 (1983).

\bibitem{Hlinka2012a}
J. Hlinka, J. Adv. Dielec. \textbf{2}, 1241006 (2012).

\bibitem{Bosak2012a}
A. Bosak, D. Chernyshov, S. Vakhrushev, and M. Krischa, Acta Cryst. \textbf{A68}, 117  (2012).

\bibitem{piezoelectricity}
S.-E. Park and T. R. Shrout, J. Appl. Phys. \textbf{82}, 1804 (1997).

\end{thebibliography}
\end{document}